\documentstyle[aps,twocolumn,epsfig,floats,prl,times,graphics]{revtex}
\tightenlines

\begin{document}

\title{Voltage-controlled wavelength conversion by terahertz electro-optic
modulation in double quantum wells}
\author {M. Y. Su, S. Carter, and M. S. Sherwin}
\address {
Physics Department and Center for Terahertz Science and Technology,
University of California, Santa Barbara, California 93106}
\author {A. Huntington and L. A. Coldren}
\address {
Materials Department, University of California,
Santa Barbara, California 93106}

\date{\today}
\maketitle
\begin{abstract}

An undoped double quantum well (DQW) was driven with a terahertz
(THz) electric field of frequency $\omega_{THz}$ polarized in the
growth direction, while simultaneously illuminated with a
near-infrared (NIR) laser at frequency $\omega_{NIR}$.  The
intensity of NIR upconverted sidebands
$\omega_{sideband}=\omega_{NIR} + \omega_{THz}$ was maximized when
a dc voltage applied in the growth direction tuned the excitonic
states into resonance with both the THz and NIR fields.  There was
no detectable upconversion far from resonance. The results
demonstrate the possibility of using gated DQW devices for
all-optical wavelength shifting between optical communication
channels separated by up to a few THz.
\end{abstract}


\narrowtext

A basic function in a wavelength division multiplexed (WDM)
optical communications network is to switch data between different
wavelength channels.  All-optical switching in a WDM network
requires the ability to shift the frequency of a near-infrared
(NIR) carrier of a NIR carrier by several terahertz
(THz)\cite{Mukherjee}.  This has been accomplished by four-wave
mixing of NIR beams in semiconductor optical
amplifiers\cite{zhou1994}, and three-wave mixing in
quasi-phase-matched AlGaAs waveguides\cite{yoo1999}.

Generation of THz optical sidebands on a NIR carrier beam has
recently been studied in various semiconductor
systems\cite{Cerne1997,Kono1997,Phillips1999}.  Great flexibility
for practical optoelectronic devices that operate at THz
frequencies lie with gated double quantum well (DQW) structures,
in which two quantum wells (QW) are separated by a thin tunnel
barrier.  Their intersubband spacings can be engineered to lie in
the THz frequency range by adjusting the width of the tunnel
barrier. Meanwhile, the optical bandgap can be separately tuned by
the width of the individual wells. Large optical nonlinearities
can be built into the DQW\cite{Capasso1994} by using an asymmetric
DQW in which carriers feel a non-centrosymmetric confinement
potential. Finally, the intersubband spacing can be tuned by
applying a DC voltage to the gates.

This paper describes experiments in which a near-infrared (NIR)
probe laser beam is modulated at THz frequencies in a gated,
asymmetric DQW. The THz field couples to an excitonic \em
intersubband \em excitation while the NIR field couples to an
excitonic \em interband \em excitation. Applying a DC voltage to
the gates can bring the intersubband transition into resonance
with the THz field, and the interband transition into resonance
with the NIR field.  When these resonance conditions are met the
NIR probe is modulated resulting in the emission of optical
sidebands which appear at frequencies
\begin{equation}
\omega_{sideband} = \omega_{NIR} + n\omega_{THz}
\end{equation}
where $\omega_{NIR}$ ($\omega_{THz}$) is the frequency of the NIR
(THz) beam and $n=\pm 1,2,\ldots$.

\begin{figure}

\begin{center}
\includegraphics[width=1.0\linewidth]{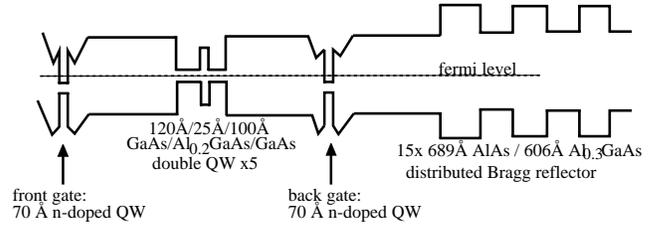}
\caption {Sample band structure, containing frontgate, active
region, backgate, and distributed Bragg reflector.  Not to scale.}
\label{band_structure}
\end{center}
\end{figure}

The sample consisted of an active region containing the DQWs, a
distributed Bragg reflector (DBR), and two gates, each consisting
of a narrow n-doped quantum-well. The band diagram of the sample
is shown in Fig. \ref{band_structure}.

The active region consisted 5 periods of DQW, each consisting of a
120 {\AA} GaAs QW and a 100 {\AA} GaAs QW separated by a 25 {\AA}
Al$_{0.2}$Ga$_{0.8}$As tunnel barrier. Each period is separated by
a 300 {\AA} Al$_{0.3}$Ga$_{0.7}$As barrier.  The active region is
thin compared to a NIR wavelength. A thicker active region
generates stronger sidebands, but is more difficult to model due
to reabsorption, phase-matching, and screening effects.

The dimensions of the DQW were designed so that the two
lowest-lying electron subbands were separated by $\approx 10$ meV
($\approx 2.4$ THz) at flat-band.  This is around the center of
the range of high-quality operating frequencies of our THz source,
the UCSB Free Electron Laser (FEL).  Within the 10 meV energy
splitting constraint, the degree of asymmetry was designed to peak
the strength of the sideband near flat-band conditions.

The DBR consisted of 15 periods of 689 {\AA}  AlAs and 606 {\AA}
Al$_{0.3}$Ga$_{0.7}$As. It had a low-temperature passband nearly
centered on the low-temperature bandgap of the DQW, making it
about 95\% reflective for the NIR probe beam and sidebands.

The active region was sandwiched between the two gates, each
consisting of a Si delta-doped 70 {\AA} QW with carrier density
$\approx 1\times 10^{12}$ cm$^{-2}$.  The gates are separated from
the active region by 3000 {\AA} Al$_{0.3}$Ga$_{0.7}$As barriers.
Since the gate QWs are much narrower than the active DQWs, the
gate QWs are transparent to both the THz and NIR beams.  A mesa
was etched and NiGeAu ohmic contacts annealed to the gate QWs.

\begin{figure}

\begin{center}
\includegraphics[width=1.0\linewidth]{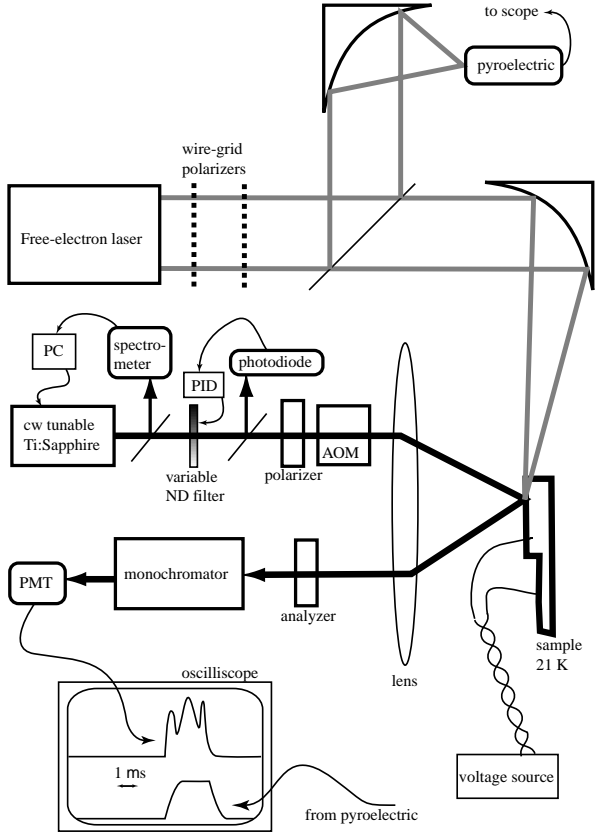}
\caption{Experimental layout. The NIR beam propagates normal to
the THz beam. The wiggles in the PMT oscilloscope trace is photon
shot noise.} \label{experiment}
\end{center}
\end{figure}

The experimental setup is illustrated in Fig. \ref{experiment}.
The sample was cooled to 21 K in a closed-cycle He crystat. The
THz beam from the FEL is polarized in the growth direction,
propagated in the QW plane and focused by a 90$^o$ off-axis
parabolic mirror (F/1) onto the cleaved edge of the sample.

NIR light from a continuous-wave Ti:sapphire laser is chopped by
an acousto-optic modulator into $\approx 25$ $\mu$s pulses which
overlap the $\approx 5$ $\mu$s FEL pulses. The
vertically-polarized NIR beam propagates normal to the THz beam,
and was focused (F/10) at a power density of $\approx$ 50 W/cm$^2$
to the same small interaction volume in the sample.  The reflected
beam, sidebands, and photoluminescence (PL) are analyzed by a
second polarizer, dispersed by a 0.85 m Raman
double-monochromator, and detected by a photomultiplier tube
(PMT).

\begin{figure}

\begin{center}
\includegraphics[width=1.0\linewidth]{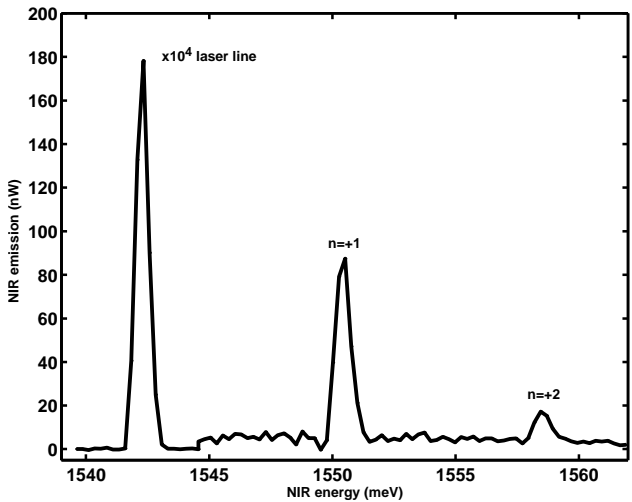}
 \caption {$n>0$ sideband spectrum at 0 V gate bias, $\hbar\omega_{NIR}= 1542$ meV,
$\hbar\omega_{THz}=8.2 meV$ (2.0 THz).}
\label{sbsplus}
\end{center}
\end{figure}

A typical $n>0$ sideband spectrum is shown in Fig. \ref{sbsplus}.
The conversion efficiency scale was calibrated by attenuating the
NIR laser to match the peak sideband signal, and factoring the
loss through the collection and dispersion optics.  We present
results for the $n=+1$ sidebands as a continuous function of gate
bias ($V_g$) and NIR laser frequency ($\omega_{NIR}$) at various
THz laser frequencies ($\omega_{THz}$).

\begin{figure}

\begin{center}
\includegraphics[width=1.0\linewidth]{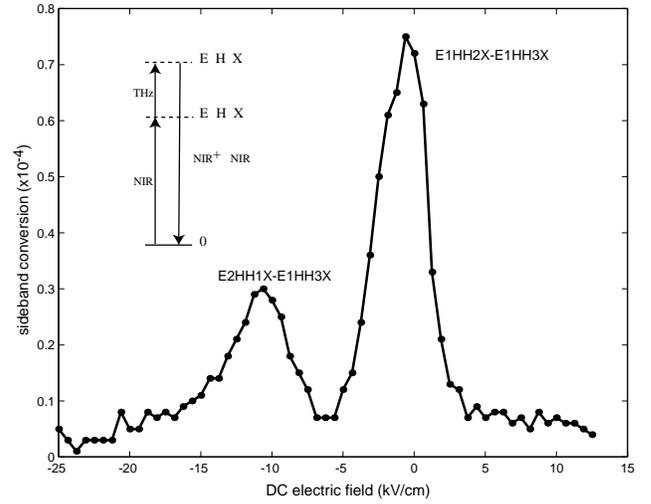}
\caption {Sideband voltage scan at $\omega_{NIR}=1546$ meV and
$\omega_{THz}=2.0$ THz (8.2 meV).  Peak assignments are derived
from a excitonic nonlinear susceptibility theory.}
\label{sbvs}
\end{center}
\end{figure}

A sideband \em voltage \em scan with $\omega_{NIR}=1546$ meV and
$\omega_{THz}=2.5$ THz is shown in Fig. \ref{sbvs}.  In this scan,
$V_g$ is scanned while all other parameters are kept constant so
that $\omega_{detect} = \omega_{NIR} + \omega_{THz}$.  Note that
at dc electric fields far from resonance there are no detectable
sidebands.  This type of scan demonstrates the voltage tunability
of the THz-NIR modulation.

The distinct peaks are labelled with excitonic transition
assignments derived from a nonlinear susceptibility calculation
for THz-NIR mixing due to excitons\cite{Su1999,SuUnpub}. The label
$E_\mu HH_\nu X$ refers to the exciton consisting of an electron
in subband $\mu$ and a heavy hole in subband $\nu$.  A
double-resonance condition holds when the NIR field is resonant
with an exciton and the THz field resonantly couples two excitons.

\begin{figure}

\begin{center}
\includegraphics[width=1.0\linewidth]{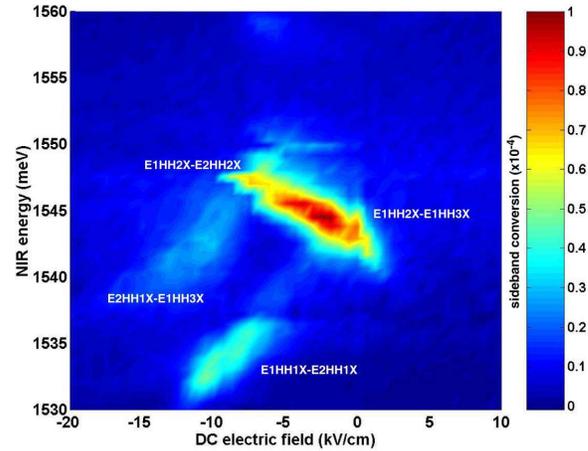}
\caption {Sideband excitation voltage scan at $\omega_{THz}=2.0$
THz (8.2 meV).}
\label{sbevs66}
\end{center}
\end{figure}

\begin{figure}

\begin{center}
\includegraphics[width=1.0\linewidth]{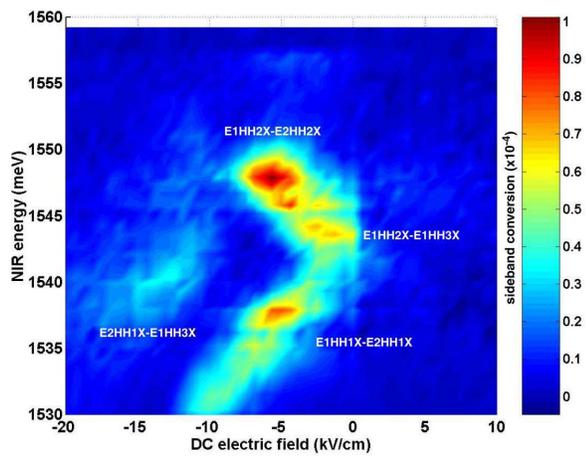}
 \caption {Sideband excitation voltage scan at $\omega_{THz}=2.5$ THz (10.4 meV).}
\label{sbevs84}
\end{center}
\end{figure}

To understand the full resonant structure of the sideband
generation process at each THz frequency, the sideband was
measured as a continuous function of both $\omega_{NIR}$ and
$V_g$.  By taking a sideband voltage scan at each $\omega_{NIR}$
we measured the sideband maps shown in Figs. \ref{sbevs66} and
\ref{sbevs84} at $\omega_{THz}=$2.0 THz and 2.5 THz (8.2 meV and
10.4 meV).  Again, peaks in the sideband emission occur when the
NIR field is resonant with an exciton and the THz field resonantly
couples two excitons; they are labelled with excitonic transition
assignments.

In summary, we have modulated a NIR laser beam at THz frequencies
by driving the excitonic intersubband transition of an asymmetric
DQW.  A pair of n-doped QW gates which are transparent to both the
NIR and THz fields allow the resonant response of the device to be
tuned by applying a dc voltage, allowing the device to act as a
voltage-controllable wavelength converter.  The peak conversion
response occurs when the NIR field is resonant with an exciton and
the THz field resonantly couples two excitons.

We gratefully thank Christoph Kadow and Art Gossard for their
input.  This research was funded by NSF-DMR 0070083.

\bibliography{/mark_su/bibtex/su_aps}
\bibliographystyle{unsrt}

\end{document}